\documentclass[aps,prd,preprint,tightenlines,superscriptaddress,
amsfonts,amssymb,amsmath,byrevtex,showpacs,nofootinbib]{revtex4-2}
\usepackage[polish,english]{babel}
\usepackage[applemac]{inputenc}
\usepackage[T1]{fontenc}
\usepackage{latexsym}
\usepackage{graphicx}
\usepackage{geometry}
\usepackage{epstopdf}
\usepackage{subfig}
\usepackage{cancel}
\usepackage{color}

\usepackage{xcolor}
\usepackage{pdfsync}
\usepackage{fancyhdr}
\usepackage{makeidx}
\usepackage[breaklinks,colorlinks,backref]{hyperref}
\hypersetup{
    colorlinks, 
    linktoc=all, 
    linkcolor=red,  
  citecolor=hyptxt,
  urlcolor=blue}
  \hypersetup{
  citebordercolor=,
  filebordercolor=green,
  linkbordercolor=blue
}
\definecolor{hyptxt}{rgb}{0.7, 0.4, 0.9}
\usepackage{bibtopic}
\newcommand{\bea}{\begin{eqnarray}}
\newcommand{\eea}{\end{eqnarray}}

\newcommand{\dR}{\mathbb R}

\newcommand{\dN}{\mathbb N}

\newcommand{\be}{\begin{equation}}
\newcommand{\ee}{\end{equation}}

\newcommand{\I}{\mathbb I}

\newcommand{\UnitOp}{\hat{1\kern-4.75pt 1}} 
\newcommand{\ket}[1]{|\kern.3ex#1\kern.3ex\rangle}
\newcommand{\bra}[1]{\langle\kern.3ex #1 \kern.3ex|}
\newcommand{\scalar}[2]{\langle\kern.3ex #1 \kern.3ex|\kern.3ex#2\kern.3ex\rangle}
\newcommand{\norm}[1]{\|\kern.3ex#1\kern.3ex \|}

\newcommand{\Group}[1]{\mathrm{#1}} 
\newcommand{\Bra}[1]{\langle #1 \vert} 
\newcommand{\Ket}[1]{\vert #1 \rangle} 
\newcommand{\BraKet}[2]{\langle #1 \vert #2 \rangle} 

\newcommand{\iu}{\texttt{i}}
\newcommand{\Res}[2]{{\rm Res}|_{#2}\left(#1\right)}

\begin{document}

\title{Quantum dynamics of gravitational massive shell}

\author{Andrzej G\'{o}\'{z}d\'{z}}
\email{andrzej.gozdz@umcs.lublin.pl}
\affiliation{Institute of Physics, Maria Curie-Sk{\l}odowska
University, pl.  Marii Curie-Sk{\l}odowskiej 1, 20-031 Lublin, Poland}

\author{Marcin Kisielowski} \email{marcin.kisielowski@gmail.com}
\affiliation{Department of Fundamental Research, National Centre for Nuclear
  Research, Pasteura 7, 02-093 Warszawa, Poland}

\author{W{\l}odzimierz Piechocki} \email{wlodzimierz.piechocki@ncbj.gov.pl}
\affiliation{Department of Fundamental Research, National Centre for Nuclear
  Research, Pasteura 7, 02-093 Warszawa, Poland}

\date{\today}

\begin{abstract}
The quantum dynamics of a self-gravitating thin matter shell in vacuum has been considered.
Quantum Hamiltonian of the system is  positive definite. Within chosen set of parameters,
the quantum shell bounces above the horizon. Considered quantum system does not collapse
to the gravitational singularity of the corresponding classical system.
\end{abstract}


\maketitle

\tableofcontents

\section{Introduction}

The dynamics of a self-gravitating thin matter shell is one of the simplest models describing gravitational
collapse of an isolated gravitational system. In the case of a spherically symmetric shell in vacuum,
a satisfactory Hamiltonian description of that dynamics has been found;  see the paper \cite{KMM} and
references therein. By shell in vacuum one means a thin matter shell with a region of flat Minkowski
space in the interior and the Schwarzschild geometry in the exterior of the shell.
The global Hamiltonian of that system is explicitly time independent and is a function
of two canonically conjugated phase space variables. That Hamiltonian is equal to the Arnowit-Deser-Misner (ADM)
mass at spacial infinity. Having well defined Hamiltonian description of matter shell, we have decided to quantize
that system to get insight into corresponding quantum dynamics. Present paper is devoted to the examination of such
an issue.

The shell system is simple enough to be treated satisfactory at classical level, and rich enough for the examination
of various aspects of corresponding quantum system. Recently, it was used for addressing the issue of the importance
of the choice of time parameter at quantum level \cite{Vaz}. It was shown that quantum theories of the shell for
different choices of time are not unitarily equivalent.

We have found that our quantum Hamiltonian is positive definite which supports its classical property.
Within chosen set of parameters describing our system, the quantum shell bounces above the horizon that is in contrast
to the classical case. It means that due to quantum effects our quantum system does not collapse to gravitational
singularity.

The paper is organized as follows:
\noindent In Sec. II we present the solution to Hamilton's dynamics restricting considerations to the subspace
of phase space for which the Hamiltonian is positive definite. In Sec. III we recall the coherent states
quantization method applied in this paper. Sec. IV concerns the calculations of the matrix elements of quantum
observables in specific basis of considered Hilbert space. That includes the operator of Hamiltonian and
operators of canonical variables. The quantum evolution of the system is presented in Sec. V. We conclude
in Sec. VI.

\noindent In the following we choose $\;G = c =1=\hbar\;$ except where otherwise noted.

\section{Classical dynamics}

For self-consistency of the present paper, we recall the main  results of Ref. \!\cite{KMM}.
Next, we present the solution to the classical dynamics.

The canonical structure of the phase space of the system ``shell+gravity'' is given by
\begin{equation}\label{Hd1}
  \omega = dp\wedge dq \, ,
\end{equation}
where $q \in \dR_+ := \{x \in \dR \;|\; x > 0\}$ is the configuration variable representing the proper volume
of the shell, and $p\in \dR$ is the momentum representing
the hyperbolic angle between the surfaces of constant time on both sides of the shell.

Hamilton's dynamics reads
\begin{eqnarray}
 \label{Hd2} \dot{q} = \frac{\partial H}{\partial p} \, , \\
 \label{Hd3} \dot{p} = - \frac{\partial H}{\partial q} \, ,
\end{eqnarray}
where the Hamiltonian is defined to be
\begin{equation}\label{Hd4}
  H (p,q):= \sqrt{\frac{q}{2}}\; \Big [ 1 - \Big( \cosh(p) - \sqrt{\frac{m^2(q)}{2q} + \sinh^2(p)} \Big)^2 \Big] \, ,
\end{equation}
and where $m(q)$ represents the total rest mass of the matter (energy) of the
shell and plays the role of the constitutive equation for the matter field of
the shell.  The dots over $q$ and $p$ in \eqref{Hd2}--\eqref{Hd3} denote
time derivatives, where the time variable is the Schwarzschild time $t$ measured
at spatial infinity. Since $H$ is time independent, the total energy of the
entire system is conserved.

Making use of \eqref{Hd2}--\eqref{Hd4} we can  determine if the shell's size increases or decreases with time.
It follows that
\begin{equation}\label{shellsize}
\dot{q}=\frac{\partial H}{\partial p} = \sinh(p) \sqrt{2q} \frac{(\cosh(p)-\sqrt{\frac{m^2}{2q}
+\sinh^2(p)})^2}{\sqrt{\frac{m^2}{2q}+\sinh^2(p)}}.
\end{equation}
Therefore the sign of $\dot{q}$ is dictated (up to singularities where the right-hand-side vanishes) by $\sinh(p)$. This means
that for positive $p$ the shell grows $\dot{q}\geq 0$ and for negative $p$ the shell shrinks $\dot{q}\leq 0$.

Any canonical transformation of the system \eqref{Hd1}--\eqref{Hd4} leads to the physically equivalent system.
The advantage of the present choice of the phase space variables is their clear physical interpretation.

The Hamiltonian function \eqref{Hd4} is equal to the total energy of considered isolated gravitational system at spatial infinity
so that it  is the ADM mass. It means, roughly speaking,  that if the density of considered matter field is
positive, the ADM mass must be positive (see, \cite{JK} and references therein). This may impose the restriction on the phase
space of considered gravitational system. The specific form of matter field may lead to the specific subspace $\Lambda$ of the
phase space $\Pi = \{(p,q)~|~p\in \dR, q \in \dR_+\}$, such that $H(p,q) > 0$ for $(p,q) \in \Lambda$. We call $\Lambda$
the  physical phase space.


Let us consider the case $m(q) := m = const$ which corresponds to the
dust matter. Since the Hamiltonian is time independent, the energy is conserved. For each value of the Schwarzschild
mass $M > 0$,   the curve
\begin{equation}\label{eq:constant_energy}
H(p,q)=M
\end{equation}
is the shell's trajectory in the phase-space.  Let us notice, that the Hamiltonian is an even function of $p$ and
therefore it is enough to look for solutions with positive momentum $p>0$. Rewriting the equation \eqref{eq:constant_energy}
(see also \eqref{Hd4}) in the form
\begin{equation}\label{cond1}
1-\sqrt{\frac{2}{q}} M = \;
 \Big( \cosh(p) - \sqrt{\frac{m^2(q)}{2q} + \sinh^2(p)} \Big)^2,
\end{equation}
we immediately notice that $q > 2 M^2$. This condition says that the proper volume $q$ is bounded from below so that
reflects the fact that
the shell is outside the event horizon of the exterior Schwarzschild solution.

The equation \eqref{eq:constant_energy} can be solved for $p$ as a function of $q$ and $M$. It can be checked by substitution
that
\begin{itemize}
\item for $M\sqrt{\frac{2}{q}} \geq \frac{m^2}{2q}$, i.e., $\sqrt{q}\geq \frac{m^2}{2\sqrt{2} M}$, the solution satisfies:
\begin{equation}\label{eq:pqI}
\cosh(p)=\frac{2-\frac{m^2}{2q}-M\sqrt{\frac{2}{q}}}{2\sqrt{1-M\sqrt{\frac{2}{q}}}}=:f(q),
\end{equation}
\item while for $M\sqrt{\frac{2}{q}} < \frac{m^2}{2q}$, i.e.,$\sqrt{q}< \frac{m^2}{2\sqrt{2} M}$, it satisfies:
\begin{equation}\label{eq:pqII}
\cosh(p)=-\frac{2-\frac{m^2}{2q}-M\sqrt{\frac{2}{q}}}{2\sqrt{1-M\sqrt{\frac{2}{q}}}}=-f(q).
\end{equation}
\end{itemize}

\begin{figure}[hbt!]
\includegraphics{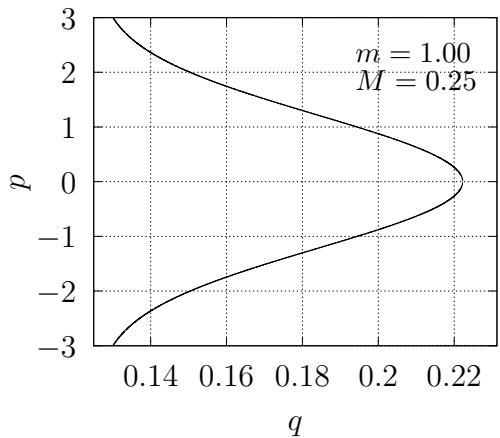}
\caption{Dust shell's trajectory with energy $H(p,q)=M<\frac{m}{2}$.}\label{fig:M_between_0_0_5_m}\label{rysunek1}
\end{figure}

At the first glance, it may seem that for $0 \leq M<\frac{m}{2}$ both relations \eqref{eq:pqI} and \eqref{eq:pqII} are
relevant and that there is a discontinuity at $q=\frac{m^2}{2\sqrt{2} M}$. However, it turns out that in this case for
$\sqrt{q}\geq \frac{m^2}{2\sqrt{2} M}$ the function $f(q)$ takes only values smaller than $1$ and therefore the only
relevant relation is \eqref{eq:pqII}. The function $-f(q)$ decreases from $+\infty$ at $2 M^2$ to $-1$ at $+\infty$.
This means that $p(q)$ decreases from $+\infty$ at $q = 2 M^2$ until $q=\frac{m^4}{8(m-M)^2}$ where $p=0$. This is
illustrated on figure \ref{fig:M_between_0_0_5_m}. The plot for $p<0$ is just a reflection by the $q$ axis.

For $M=\frac{m}{2}$ the function $f(q)$ takes only values smaller than $1$ and therefore there is no solution.

\begin{figure}[!tbp]
  \centering
  \subfloat[$\frac{m}{2}<M<m$]{\includegraphics[scale=0.75]{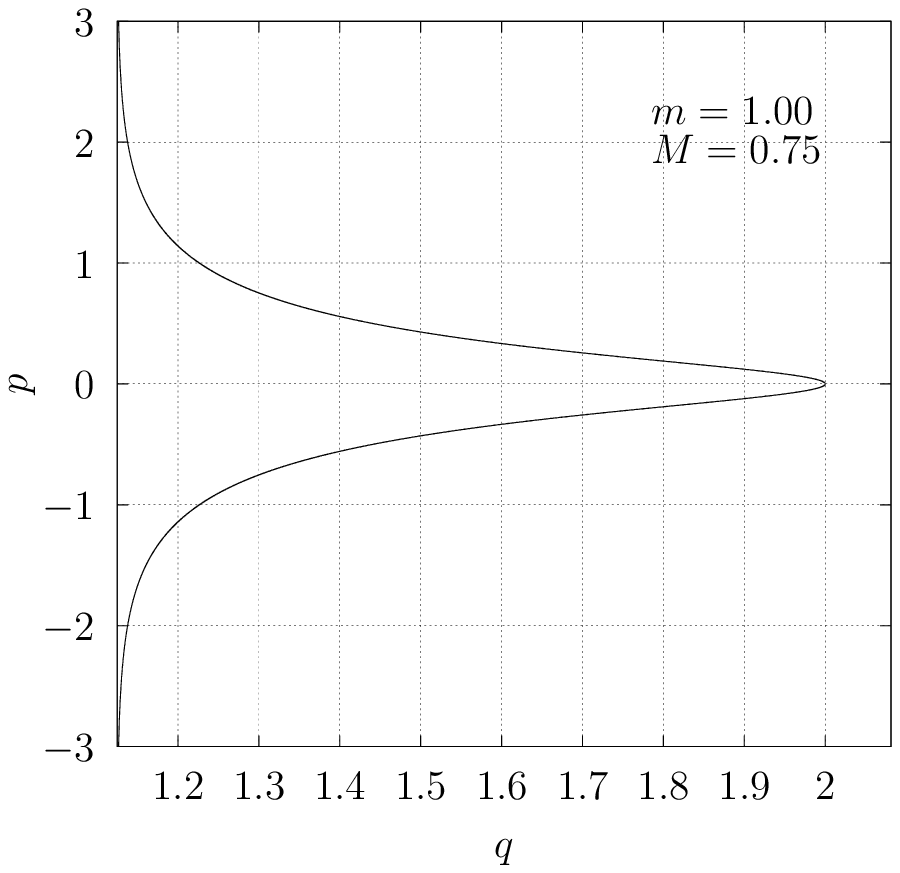}\label{fig:M_between_0_5_m_m}}
  \hfill
  \subfloat[$M\geq m$]{\includegraphics[scale=0.75]{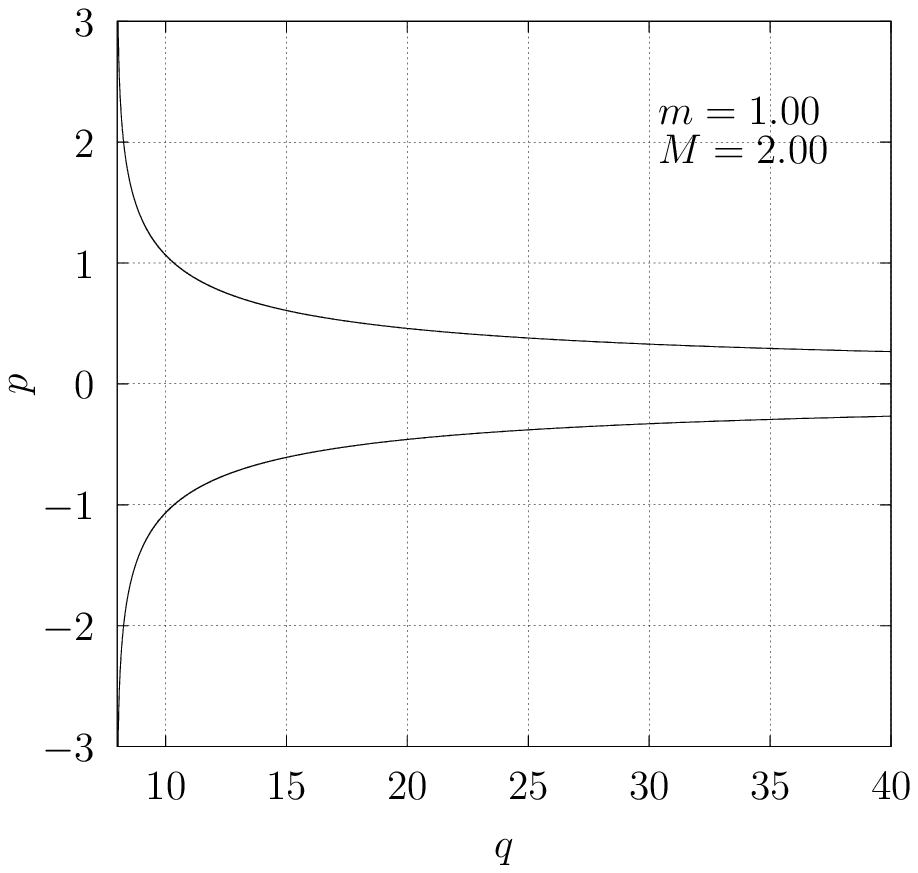}\label{fig:M_between_m_infty}}
  \caption{Dust shells' trajectories with energy $H(p,q)=M>\frac{m}{2}$.}
\end{figure}

If $M>\frac{m}{2}$, we have  $\sqrt{q}< \frac{m^2}{2\sqrt{2} M}$  so the function $f(q)$ is fully determined by
the relation \eqref{eq:pqI}. The case $M>\frac{m}{2}$ splits further into two other cases:
\begin{enumerate}
\item $\frac{m}{2}<M<m$: In this case the function $f(q)$ decreases from $+\infty$ at $q = 2 M^2$, reaches
a minimal value smaller than $1$ and grows asymptotically to $1$  as $q$ goes to $+\infty$. This means that
$p(q)$ decreases from $+\infty$ at $q = 2M^2$ until $q=\frac{m^4}{8(m-M)^2}$ where $p=0$.
This is illustrated on figure \ref{fig:M_between_0_5_m_m}. The plot for $p<0$ is just a reflection by the $q$ axis.
\item $M\geq m$: In this case the function $f(q)$ decreases from $+\infty$ at $q = 2 M^2$ to $1$ at $+\infty$. This means that
that $p(q)$ decreases from $+\infty$ at $q = 2 M^2$ to $0$ at $q \rightarrow +\infty$. This is illustrated on figure
\ref{fig:M_between_m_infty}.
\end{enumerate}

Let us examine the issue of the positive definiteness of $H(p,q)$.  For $p = 0$, Eq. \eqref{Hd4}
reads $ H(0,q) = m - m^2 / \sqrt{8q}~$ so that we have
\begin{equation}\label{posdef}
  H(0,q) < 0~~~\mbox{for}~~~q < m^2/8.
\end{equation}
Since wee need to have $H(p,q) > 0$, the physical phase space $\Lambda$ should be a subspace of the entire phase space $\Pi$.
Some regions of $\Pi$, for instance defined by \eqref{posdef}, may lead to the breaking of this condition.

\section{Quantum level}

The affine coherent states quantization applied in this article is based on the
formalism presented in our papers \cite{Gozdz:2018aai} and \cite{AWT}.

The physical phase space of our gravitational system
\begin{equation}\label{Q1}
  \Pi := \{(p,q)\;|\; p\in \dR, q \in \dR_+\} \, ,
\end{equation}
can be identified with the affine group $\Group{G}
\equiv \Group{Aff}(\dR)$, by defining the multiplication law as follows
\begin{equation}\label{Q2}
  (p^\prime, q^\prime)\cdot (p,q):= (q^\prime p + p^\prime, q^\prime q)\, ,
\end{equation}
with the unity $(0,1)$ and the inverse
\begin{equation}\label{Q3}
(p^\prime, q^\prime)^{-1} =
(-\frac{p^\prime}{q^\prime}, \frac{1}{q^\prime}) \, .
\end{equation}

\subsection{Affine coherent states}

The affine group has two, nontrivial, inequivalent irreducible unitary
representations \cite{Gel,AK1,AK2}.  Both are realized in the Hilbert space
$\mathcal{H}=L^2(\dR_+, d\nu(x))$, where $d\nu(x):=dx/x$ is an invariant
measure on the multiplicative group $(\dR_+,\cdot)$.
In what follows we choose the one  defined by the following action
\begin{equation}\label{Q4}
U(p,q)\psi(x)= e^{i p x} \psi(qx)\, ,
\end{equation}
where\footnote{We use Dirac's notation whenever we wish to deal with abstract
vector, instead of functional representation of the vector.}  $\psi(x) = \langle x|\psi\rangle$ and $|\psi\rangle
\in L^2(\dR_+, d\nu(x))$.

We define the integrals over the affine group
$\Group{G} =\Group{Aff}(\dR)$ as follows
\begin{equation}\label{Q5}
\int_{\Group{G}} d\mu(p,q)= \frac{1}{2\pi}
\int_{-\infty}^{+\infty} dp \int_{0}^\infty \frac{dq}{q^2} \, .
\end{equation}

Fixing the normalized vector $\Ket{\Phi} \in L^2(\dR_+, d\nu(x))$, called the
fiducial vector, we can define a continuous family of  affine
coherent states $\Ket{p,q} \in L^2(\dR_+, d\nu(x))$ as follows
\begin{equation}\label{Q6}
\Ket{p,q} = U(p,q) \Ket{\Phi}.
\end{equation}

The  irreducibility of the representation, used to define the coherent
states \eqref{Q6}, enables making use of Schur's lemma \cite{BR}, which leads
to the resolution of the unity in $L^2(\dR_+, d\nu(x))$
\begin{equation}\label{Q7}
\frac{1}{ A_\Phi}\int_{\Group{G}}  d\mu(p,q) \Ket{p,q}\Bra{p,q} = \I \; ,
\end{equation}
where the constant $A_\Phi$ can be determined by using any arbitrary, normalized
vector $\Ket{f} \in L^2(\dR_+, d\nu(x))$ as follows
\begin{equation}\label{im3b}
A_\Phi = \int_{\Group{G}}d\mu(p,q)\,\BraKet{f}{p,q}\BraKet{p,q}{f} = \int_0^\infty \frac{dq}{q^2} |\Phi(q)|^2 \, .
\end{equation}

\subsection{Quantum observables}

Using the resolution of the identity \eqref{Q7}, we define the
quantization of a classical observable $f$ as follows \cite{Ber}
\begin{equation}\label{im8}
  \mathcal{F} \ni f \longrightarrow  \hat{f} :=
\frac{1}{A_\Phi}\int_{\Group{G}} d\mu(p,q)
\Ket{p,q} f(p,q) \Bra{p,q}  \in \mathcal{A} \, ,
\end{equation}
where $\mathcal{F}$ is a vector space of real continuous functions on a phase
space, and $\mathcal{A}$ is a vector space of operators (quantum observables)
acting in the Hilbert space $L^2(\dR_+, d\nu(x))$. It is clear that \eqref{im8}
defines a linear mapping and the observable $\hat{f}$ is a symmetric
operator.  Self-adjointness of $\hat{f}$ is an open problem as symmetricity
  does not assure self-adjointness so that further examination is required
\cite{Reed}.

In Appendix~\ref{basis} we define an orthonormal basis (for any fixed value of the
parameter $\alpha > -1$) of the
unitary irreducible representation of considered affine group. This basis can be
used in concrete calculations. To make these calculations feasible, we use the
technics of generating functions for generalized Laguerre polynomials
$L_n^{(\alpha)}$ (see, e.g.  \cite{Arfken2011}).  For this purpose, it is
convenient to restrict the upper label $\alpha$ of the functions
$L_n^{(\alpha)}(x)$ to any fixed integer. In this case, the generating function for the
Laguerre polynomials reads:
\begin{equation}
\label{eq:GenFunLaggPolyn}
\frac{\exp(-\frac{xz}{1-z})}{(1-z)^{\alpha+1}}=
\sum _{n =0}^\infty L_n^{(\alpha)}(x) z^n \quad \mbox{if } |z| <1 \ .
\end{equation}
For calculation of the matrix elements of the operator (\ref{im8}), in the basis defined in
App. \!\ref{basis}, one needs to calculate the overlaps between the coherent state vectors
and the vectors of this orthonormal basis. To perform these calculations explicitly, we
choose the fiducial  vector for our coherent states as follows
\begin{equation}
\label{eq:FiducialVector}
 \Phi^{(\alpha)}(x) := e^{(\alpha)}_0 (x)= \sqrt{\frac{1}{\alpha!}}
x^{\frac{1+\alpha}{2}} e^{-\frac{x}{2}} \, .
\end{equation}
In this case one gets
\begin{equation}
\label{eq:ScalarProdepq}
\BraKet{e^{(\alpha)}_n (x)}{pq}
= \sqrt{\frac{n!}{(n+\alpha)! \alpha!}}
q^{\frac{1+\alpha}{2}} \int_0^\infty \frac{dx}{x} e^{ipx} x^{1+\alpha}
e^{-\frac{q+1}{2}x} L_n^{(\alpha)}(x) \, .
\end{equation}
Let us define the function
\begin{eqnarray}
\label{eq:epqFunction}
&& \mathrm{epq}(\alpha,p,q;z) :=
\sum_n \BraKet{e^{(\alpha)}_n (x)}{pq}
\left[\sqrt{\frac{n!}{(n+\alpha)! \alpha!}}\right]^{-\frac{1}{2}} z^n
\nonumber \\
&& =\left(\frac{\sqrt{q}}{1-z}\right)^{1+\alpha}
\int_0^\infty dx\, x^\alpha
\exp\left(-\left[\frac{q+1}{2}+\frac{z}{1-z}-ip \right]x \right) \, .
\end{eqnarray}
The required overlaps turn out to be defined by the derivatives of these functions calculated at $z=0$ as follows
\begin{equation}
\label{eq:ScalarProdepq2}
\BraKet{e^{(\alpha)}_n (x)}{pq}
= \frac{1}{\sqrt{\alpha! n! (n+\alpha)!}}
\left[ \frac{d^n}{dz^n} \mathrm{epq}(\alpha,p,q;z)  \right]_{|z=0} \, .
\end{equation}
Using exactly the same method, the matrix elements of the operator (\ref{im8})
are found to be
%
\begin{eqnarray}
\label{eq:MatrElemOpf}
&& \Bra{e^{(\alpha)}_n }\hat{f}\Ket{e^{(\alpha)}_{n'} }=
\frac{1}{A_\Phi}
\frac{1}{\sqrt{\alpha! n! (n+\alpha)! \alpha! {n'}! (n'+\alpha)!} }
\nonumber \\
&&\cdot\left[ \frac{\partial^n}{\partial z_1^n} \frac{\partial^{n'}}{\partial z_2^{n'}}
\int_{\Group{Aff}(\dR)} d\mu_L(p,q)
\mathrm{epq}(\alpha,p,q;z_1)f(p,q) \mathrm{epq}(\alpha,p,q;z_2)^\star
\right]_{|z_1=z_2=0} \, .
\end{eqnarray}
The simplest basis, satisfying required conditions, is obtained by taking $\alpha=1$.  In
this case, the function $\mathrm{epq}(1,p,q;z)$ reads:
\begin{equation}\label{eq:epqFunAlpha1}
\mathrm{epq}(1,p,q;z)=\frac{q}{(1-z)^2}
\left( \frac{q+1}{2}+ \frac{z}{1-z}-ip \right)^{-2} \, .
\end{equation}
%

\subsection{Quantum dynamics}

The mapping \eqref{im8} applied to the classical Hamiltonian \eqref{Hd4} reads
\begin{equation}\label{Qd1}
 \hat{H}_{\rm unbounded}= \frac{1}{A_{\Phi} }\int_{G}  d\mu(p,q)
  \Ket{p,q}H(p,q)\Bra{p,q}\, .
\end{equation}
However, our classical analysis applies only to the region of phase space for which $H(p,q)>0$.

An important problem is introducing  constraints into the integral
quantization approach.
This quantization is based on deformation of quantum measure represented by a set of positive
self-adjoint operators determining the operator valued measure (POV). They are considered as
generalization of more standard self-adjoint quantum observables \cite{Busch1996}.

In our case the set of operators
\begin{equation}
\label{POVM}
\hat{M}(Q):= \frac{1}{A_{\phi}} \int_G d\mu(p,q) \Ket{p,q}\chi_Q (g) \Bra{p,q}\, ,
\end{equation}
where $\chi_Q (g) =1$ if $g \in Q$ and $0$ otherwise, and  where $Q \subset \Group{G} =\Group{Aff}(\dR)$,
describe the localization of the system in
the subspace $Q$ of the phase space $\Group{G}$.

The normalization condition 
\begin{equation}
\label{POVM1}
\hat{M}(G)= \frac{1}{A_{\phi}} \int_G d\mu(p,q) \Ket{p,q} \Bra{p,q}= \UnitOp \,
\end{equation}

is required to get the so called minimal
probabilistic interpretation of quantum mechanics \cite{Busch1996}:
\begin{equation}
\label{POVM2}
\mathrm{Prob}(Q,\Psi)= \Bra{\Psi}\hat{M}(\Group{Q}) \ket{\Psi} \, ,
\end{equation}
which describes the probability of finding our system in $Q$, under condition
that this system is in the state $\Psi$.

On the other hand, to construct the condition (\ref{POVM1}) one
needs to integrate over the whole group manifold and the representation
(\ref{Q6}) has to be irreducible.
This excludes  the possibility of using a smaller region $\Omega \subset \Group{G}$
of the phase space to fulfill the classical constraint, where
$\Omega=\{(p,q): H(p,q)>0\}$.

To have a consistent quantization method, the only possibility of quantizing any
observable restricted to a smaller region of the phase space is to quantize it
over the whole phase space, represented in this approach by the group
$\Group{G}$.

Following these requirements, to keep physical interpretation of the classical
Hamiltonian, we quantize it as a function restricted to the required region of the
phase space by considering:
\begin{equation}\label{Qd1}
 \hat{H}= \frac{1}{A_{\Phi} }\int_{G}  d\mu(p,q)
  \Ket{p,q}\theta(H(p,q))H(p,q)\Bra{p,q}\, ,
  \end{equation}
where $\theta$ is the Heaviside theta function.

An important feature of the operator $\hat{H}$ is that it acts in a
nontrivial way on the whole phase space. Let $(p',q') \not\in \Omega$, then
\begin{equation}
\label{ActRestrH}
 \hat{H} \Ket{p',q'} = \frac{1}{A_{\Phi} }\int_{\Omega}  d\mu(p,q)
  \Ket{p,q}\theta(H(p,q))H(p,q)\BraKet{p,q}{p',q'}
\end{equation}
is usually a non-zero vector.
This behavior is due to non-orthogonality of the states $\Ket{p,q}$, i.e., the
quantum phase space regions $\Omega$ and $\Group{G} \setminus \Omega$ are not
independent.

 Another very important problem related  is the quantization
of the elementary observables $(p,q)$.  These observables
represent position of the system in the quantum full phase space
$\Group{G}$. The corresponding quantum operators $(\hat{p},\hat{q})$  should satisfy
the following consistency conditions \cite{AOW}:
\begin{equation}\label{cond}
\Bra{p,q}\hat{p}\Ket{p,q}=p \text{~~~and~~~~}  \Bra{p,q}\hat{q}\Ket{p,q}=q  \, ,
\end{equation}
where
\begin{equation}\label{pandq}
  \hat{p} = \frac{1}{A_{\Phi} }\int_{G}  d\mu(p,q)\Ket{p,q} p \Bra{p,q} \text{~~~and~~~~}  \hat{q} = \frac{1}{A_{\Phi} }\int_{G}  d\mu(p,q)
  \Ket{p,q} q \Bra{p,q} \, .
\end{equation}
Satisfying \eqref{cond} is possible only when the operators are defined
on the entitre phase space $\Group{G}$.
The consistency conditions support the physical interpretation of the POV
measure (\ref{POVM}).

The above analysis allow us to be consistent and to quantize the restricted
form of the Hamiltonian $H$ and unrestricted form of the elementary
observables $(p,q)$.

We will restrict to this region of a phase space by considering:
\begin{equation}\label{Qd1}
 \hat{H}= \frac{1}{A_{\Phi} }\int_{G}  d\mu(p,q)
  \Ket{p,q}\theta(H(p,q))H(p,q)\Bra{p,q}\, ,
\end{equation}
where $\theta$ is the Heaviside theta function. Let us notice that the operator $\hat{H}$ is positive definite. Indeed, for any state $\ket{\Psi}$ we have:
\begin{equation}
 \scalar{\Psi}{\hat{H}\Psi}= \frac{1}{A_{\Phi} }\int_{G}  d\mu(p,q)
  |\Psi(p,q)|^2\theta(H(p,q))H(p,q)\, >0.
\end{equation}

The quantum evolution of our gravitational system is defined by the
Schr\"{o}dinger equation:
\begin{equation}\label{Qd2}
  i  \frac{\partial}{\partial s}|\Psi (s) \rangle
= \hat{H} |\Psi (s) \rangle \; ,
\end{equation}
where $|\Psi \rangle \in L^2(\dR_+, d\nu(x))$, and where $s$ is an evolution
parameter of the quantum level.

In general, the parameters $t$ of the classical level and $s$ are different. To get the consistency
between the classical and quantum levels we postulate that $t = s$, which defines the  time
variable at both levels.  This way we support the interpretation that Hamiltonian is the generator
of classical and corresponding quantum dynamics.

\section{Matrix elements of quantum observables}

In this section we  calculate the matrix elements for the operators $\hat{H}, \hat{q}$, and $\hat{p}$.
The matrix elements of $\hat{H}$ are calculated numerically using \eqref{eq:Hamiltonian_matrix}.
We  calculate the matrix elements of $\hat{q}$ and $\hat{p}$ analytically. The computations are based
on a new formula for the basis elements $e^{(1)}_m$ that we derive in the first subsection. The results
of this section are used in Sec. \!\ref{sc:evolution_of_obsevables}  to find an evolution of the quantum
observables.

\subsection{New expression for the basis elements}

In what follows, we use the basis with $\alpha=1$. Since $\frac{z}{1-z}=\frac{1}{1-z}-1$
and $\frac{d}{dz}\frac{1}{1-z} = \frac{1}{(1-z)^2} $, we can simplify the expression for the basis element:
\begin{equation}\label{eq:basis}
\BraKet{e^{(1)}_n}{pq}=-\frac{q \sqrt{n+1}}{ (n+1)!}\left[ \frac{d^{n+1}}{dz^{n+1}} \int_0^\infty dx\,
\exp\left(-\left[\frac{q-1}{2}+\frac{1}{1-z}-ip \right]x\right) \right]_{|z=0}
\end{equation}
The technical problem is now to calculate $\frac{d^{n}}{dz^{n}} f(z)$, where $f(z)$ is a composite function:
\begin{equation}
f(z)=F(G(z)) = \exp\left(- G(z) x\right), {\rm\ where\ }G(z)=\frac{1}{1-z}.
\end{equation}
We will use the Faa di Bruno's formula \cite{FaaDiBruno, BellPolynomialEncyclopedia}:
\begin{equation}\label{eq:FaaDiBruno}
\frac{d^n}{dz^n} F(G(z))=\sum_{k=1}^n F^{(k)}(G(z)) \cdot B_{n,k}(G'(z),G''(z),\ldots, G^{(n-k+1)}(z)),
\end{equation}
where $B_{n,k}(x_1,\ldots, x_{n-k+1})$ are Bell polynomials \cite{BellPolynomial,BellPolynomialEncyclopedia}.
In the formula above we denote by $F^{(k)}$ the k-th derivative of $F$.
The Bell polynomial $B_{n,k}(z_1,z_2,\ldots, z_{n-k+1})$ is given by
\begin{equation}
B_{n,k}(z_1,z_2,\ldots, z_{n-k+1})=\sum \frac{n!}{j_1! j_2! \ldots j_{n-k+1}!}
\left(\frac{z_1}{1!}\right)^{j_1} \left(\frac{z_2}{2!}\right)^{j_2} \ldots \left(\frac{z_{n-k+1}}{(n-k+1)!}\right)^{j_{n-k+1}},
\end{equation}
where the sum is over all sequences $j_1,\ldots, j_{n-k+1} \in \mathbb{N}$ such that
\begin{eqnarray}
&&j_1+j_2+\ldots+j_{n-k+1} =k,\\
&&j_1 + 2\, j_2 + 3\,j_3+\ldots+(n-k+1)j_{n-k+1}=n.
\end{eqnarray}
We are interested only in derivatives at $z=0$, therefore in equation \eqref{eq:FaaDiBruno} we should put $z=0$.
Let us notice that
\begin{equation}
G(0)=1, \quad G^{m}(0)=m!\, .
\end{equation}
Therefore
\begin{equation}\label{eq:FaaDiBrunoEvaluated}
\frac{d^n}{dz^n} F(G(0))=\sum_{k=1}^n (-1)^k x^k e^{-x}\,B_{n,k}(1!,2!,\ldots, (n-k+1)!).
\end{equation}
The coefficient $B_{n,k}(1!,2!,\ldots, (n-k+1)!)$ can be expressed in terms of the Lah number \cite{BellPolynomialEncyclopedia,Lah}:
\begin{equation}
B_{n,k}(1!,2!,\ldots, (n-k+1)!)=|L_{n,k}|,
\end{equation}
where
\begin{eqnarray}
L_{n,k}=(-1)^n \binom{n-1}{k-1}\frac{n!}{k!},\quad n\geq k \geq 1,\\
L_{0,0}=1,\quad L_{n,0}=0,\,n\geq 1.
\end{eqnarray}
We extend the sum in \eqref{eq:FaaDiBrunoEvaluated} to $k=0$ and obtain a formula valid for $n=0$:
\begin{equation}
\frac{d^n}{dz^n} F(G(0))=\sum_{k=0}^n  (-1)^k |L_{n,k}|\, x^k e^{-x}.
\end{equation}

In order to calculate $\BraKet{e^{(1)}_n}{pq}$, we need to evaluate the integral
\begin{equation}
\int_0^{\infty} dx\ x^{k} \exp\left(-\left(\frac{q+1}{2}-\iu p\right)x\right) = \frac{ k!}{\left(\frac{q+1}{2} -\iu p \right)^{k+1}}.
\end{equation}
This integral directly follows from the Euler's integral formula (see for example equation 6.1.1 from \cite{AbramowitzStegun}:
\begin{equation}
\Gamma(n)=z^n \int_0^\infty dx\, x^{n-1} e^{-z x},~\quad {\rm for\ } ~\Re(n)>0,~ \Re{(z)}>0.
\end{equation}
In the formula above $\Gamma(z)$ is the Gamma function.
Inserting this evaluation into \eqref{eq:basis} gives:
\begin{equation}\label{eq:basis_calculated}
\BraKet{e^{(1)}_n}{pq}= \sqrt{n+1} \sum_{k=0}^{n+1} \frac{(-1)^{k+1} k!}{ (n+1)!} |L_{n+1,k}| \frac{q}{\left(\frac{q+1}{2} -\iu p \right)^{k+1}} \, .
\end{equation}
We will define coefficients $E_{n,k}$:

\begin{equation}
E_{n,k}=\frac{(-1)^{k+1}  k!}{n!} |L_{n,k}|.
\end{equation}
Due to cancellation of terms, the coefficients have the following explicit form:
\begin{eqnarray}
E_{n,k}= (-1)^{k-1} \binom{n-1}{k-1},\quad n\geq k \geq 1,\\
E_{0,0}=\frac{(-1)^{k-1} k!}{n! } ,\quad E_{n,0}=0,\,n\geq 1.
\end{eqnarray}
Let us notice, that in our formulas we need only the coefficients with $n\geq 1$. Since, $E_{n,0}=0$ for $n\geq 1$,
we can assume that $k\geq 1$. The formula \eqref{eq:basis_calculated} takes now a more compact form:
\begin{equation}
\BraKet{e^{(1)}_n}{pq}= \sqrt{n+1}\sum_{k=1}^{n+1} (-1)^{k-1} \binom{n}{k-1}
\frac{q}{\left(\frac{q+1}{2} -\iu p \right)^{k+1}} \, .
\end{equation}
We will change the summation variable into $k'=k-1$ and obtain:

\begin{equation}
\BraKet{e^{(1)}_n}{pq}=\frac{q\,\sqrt{n+1}}{ \left(\frac{q+1}{2} -\iu p \right)^{2}}\,\sum_{k=0}^{n} \binom{n}{k}
\frac{(-1)^{k}}{\left(\frac{q+1}{2} -\iu p \right)^{k}}.
\end{equation}
The last sum is the binomial expansion. Therefore
\begin{equation}\label{eq:basis_final_form}
\BraKet{e^{(1)}_n}{pq}=\frac{q\,\sqrt{n+1}}{\left(\frac{q+1}{2} -\iu p \right)^{2}}\,
\left( 1 - \frac{1}{\frac{q+1}{2} -\iu p }\right)^n=q\,\sqrt{n+1}\, \frac{\left(\frac{q-1}{2} -\iu p\right)^n}
{\left(\frac{q+1}{2} -\iu p\right)^{n+2}}.
\end{equation}

\subsection{The matrix elements of the Hamiltonian operator}

The new expression for the basis elements that we found in the previous section allows us to calculate the matrix
elements of the Hamiltonian:
\begin{equation}
H_{nm}:=\BraKet{e^{(1)}_n}{\hat{H} e^{(1)}_m}=\frac{1}{A_{\Phi}} \int_G d\mu(p,q) \BraKet{e^{(1)}_n}{pq}
\theta(H(q,p)) H(p,q) \BraKet{e^{(1)}_m}{pq}^*.
\end{equation}
After applying the formula \eqref{eq:basis_final_form} to the equation above, we obtain:
\begin{equation}\label{eq:Hamiltonian_matrix}
H_{nm}:=\frac{\sqrt{(n+1)(m+1)}}{2\pi }\,\int_0^{\infty} dq \int_{-\infty}^{+\infty}
dp\ \theta(H(q,p))H(q,p) \frac{\left(\frac{q-1}{2} -\iu p \right)^{n}\left(\frac{q-1}{2}
+\iu p \right)^{m}}{\left(\frac{q+1}{2} -\iu p \right)^{n+2}\left(\frac{q+1}{2} +\iu p \right)^{m+2}} .
\end{equation}
Since $\hat{H}$ is hermitian, it is enough to calculate the lower triangular part of the matrix,
i.e. $H_{nm}$ for $n\geq m$. In this case:
\begin{equation}
H_{n m}=\frac{\sqrt{(n+1)(m+1)}}{2\pi }\int_0^{\infty} dq \int_{-\infty}^{+\infty} dp\ \theta(H(q,p))H(q,p)
\frac{\left(\frac{q-1}{2} -\iu p \right)^{n}\left(\frac{q-1}{2} +\iu p \right)^{m}\left(\frac{q+1}{2}
+\iu p \right)^{n-m}}{\left(\left(\frac{q+1}{2}\right)^2 + p^2 \right)^{n+2}}.
\end{equation}
In order to calculate the real and the imaginary part of the matrix, we write the Hamiltonian in the form
\begin{equation}\label{ab1}
H_{n m}=\frac{\sqrt{(n+1)(m+1)}}{2\pi }\int_0^{\infty} dq \int_{-\infty}^{+\infty} dp\ \theta(H(q,p))H(q,p)
\frac{\left(\left(\frac{q-1}{2}\right)^2+p^2 \right)^m \left( \frac{q^2-1}{4} + p^2-\iu p\right)^{n-m}}
{\left(\left(\frac{q+1}{2}\right)^2 + p^2 \right)^{n+2}}.
\end{equation}
After expanding the expression $\left( \frac{q^2-1}{4} + p^2-\iu p\right)^{n-m}$ in powers of $\iu$,
we notice that the real part of the integrant is an even function in $p$ and the imaginary part is
an odd function in $p$. As a result, the imaginary part of $H_{nm}$ vanishes:
\begin{equation}
\Im H_{nm}=0.
\end{equation}
We conclude that $H_{nm}$ is a real symmetric matrix. We will calculate the matrix elements by evaluating
the integrals over $q$ and $p$ numerically.

The diagonal elements $H_{nn}$ are positive  since the integrand of \eqref{ab1} is a positive function
of the variables $q$ and $p$.

\subsection{The matrix elements of the momentum operator}
The matrix elements of the momentum operator:
\begin{equation}
p_{nm}:=\frac{\sqrt{(n+1)(m+1)}}{2\pi }\,\int_0^{\infty} dq \int_{-\infty}^{+\infty} dp\ \frac{p \left(\frac{q-1}{2} -\iu p \right)^{n}\left(\frac{q-1}{2} +\iu p \right)^{m}}{\left(\frac{q+1}{2} -\iu p \right)^{n+2}\left(\frac{q+1}{2} +\iu p \right)^{m+2}} .
\end{equation}
The integral over $p$ can be done using a contour method. We take the sunset contour in the lower half plane and obtain
\begin{equation}
p_{nm}:=-\iu \sqrt{(n+1)(m+1)}\,\int_0^{\infty} dq\ \Res{\frac{p \left(\frac{q-1}{2} -\iu p \right)^{n}\left(\frac{q-1}{2} +\iu p \right)^{m}}{\left(\frac{q+1}{2} -\iu p \right)^{n+2}\left(\frac{q+1}{2} +\iu p \right)^{m+2}}}{p=-\iu \frac{q+1}{2}} .
\end{equation}
We will calculate now the residuum:
\begin{multline}
\Res{\frac{p \left(\frac{q-1}{2} -\iu p \right)^{n}\left(\frac{q-1}{2} +\iu p \right)^{m}}{\left(\frac{q+1}{2} -\iu p \right)^{n+2}\left(\frac{q+1}{2} +\iu p \right)^{m+2}}}{p=-\iu \frac{q+1}{2}}=\\=\iu^{n+2} \frac{1}{(n+1)!} \frac{d^{n+1}}{d p^{n+1}} \left( {\frac{p \left(\frac{q-1}{2} -\iu p \right)^{n}\left(\frac{q-1}{2} +\iu p \right)^{m}}{\left(\frac{q+1}{2} +\iu p \right)^{m+2}}}\right)=\\= \frac{
\iu^{n+2}}{(n+1)!} p \frac{d^{n+1}}{d p^{n+1}} \left( {\frac{ \left(\frac{q-1}{2} -\iu p \right)^{n}\left(\frac{q-1}{2} +\iu p \right)^{m}}{\left(\frac{q+1}{2} +\iu p \right)^{m+2}}}\right)|_{p=-\iu \frac{q+1}{2}}+\\+\frac{\iu^{n+2}}{n!} \frac{d^{n}}{d p^{n}} \left( {\frac{ \left(\frac{q-1}{2} -\iu p \right)^{n}\left(\frac{q-1}{2} +\iu p \right)^{m}}{\left(\frac{q+1}{2} +\iu p \right)^{m+2}}}\right)|_{p=-\iu \frac{q+1}{2}}.
\end{multline}
Let us notice that the two terms in the last line are proportional to expressions of the form:
\begin{equation}\label{eq:I_definition}
I(l,n,m,q)=\frac{1}{l!} \frac{d^{l}}{d p^{l}} \left( {\frac{ \left(\frac{q-1}{2} -\iu p \right)^{n}\left(\frac{q-1}{2} +\iu p \right)^{m}}{\left(\frac{q+1}{2} +\iu p \right)^{m+2}}}\right)|_{p=-\iu \frac{q+1}{2}}
\end{equation}
with $l=n$ or $l=n+1$. We will expand the derivative using a Leibnitz rule generalized to three factors:
\begin{multline}\label{eq:I_def}
I(l,n,m,q)=\frac{1}{l!} \sum_{\substack{k_1, k_2, k_3 \\ k_1+k_2+k_3=l}} \binom{l}{k_1,k_2,k_3} \frac{d^{k_1}}{d p^{k_1}} \left(\frac{q-1}{2} -\iu p \right)^{n} \cdot \\ \cdot \frac{d^{k_2}}{d p^{k_2}} \left(\frac{q-1}{2} +\iu p \right)^{m} \frac{d^{k_3}}{d p^{k_3}} \left(\frac{q+1}{2} +\iu p \right)^{-m-2}|_{p=-\iu \frac{q+1}{2}},
\end{multline}
where $\binom{l}{k_1,k_2,k_3}=\frac{l!}{k_1! k_2! k_3!}$. The derivatives of power functions can be calculated and give ($j,k\in \mathbb{Z}, j\geq 0 ,\ k \geq 0$):
\begin{eqnarray}
\frac{d^k}{d x^k} x^j=\begin{cases}
\frac{j!}{(j-k)!}\cdot x^{j-k},& {\rm for}\ j\geq k,\\
0, & {\rm for}\ j<k,
\end{cases}\\
\frac{d^k}{d x^k} x^{-j}=(-1)^{k}\frac{(j+k-1)!}{(j-1)!}\cdot x^{-j-k}.
\end{eqnarray}
Inserting this result into \eqref{eq:I_def} gives:
\begin{multline}
I(l,n,m,q)=\frac{1}{l!} \sum_{\substack{k_1, k_2, k_3 \\ k_1+k_2+k_3=l \\ k_1 \leq n, k_2 \leq m}} \binom{l}{k_1,k_2,k_3} \frac{n!}{(n-k_1)!} (-\iu)^{k_1} \left(\frac{q-1}{2} -\iu p \right)^{n-k_1} \cdot \\ \cdot (\iu)^{k_2} \frac{m!}{(m-k_2)!} \left(\frac{q-1}{2} +\iu p \right)^{m-k_2} (-\iu)^{k_3} \frac{(m+1+k_3)!}{(m+1)!} \left(\frac{q+1}{2} +\iu p \right)^{-m-2-k_3}|_{p=-\iu \frac{q+1}{2}}.
\end{multline}
After evaluating at $p=-\iu \frac{q+1}{2}$ we get:
\begin{multline}\label{eq:I_sum}
I(l,n,m,q)=(-1)^n(\iu)^l \sum_{\substack{k_1, k_2, k_3 \\ k_1+k_2+k_3=l \\ k_1 \leq n, k_2 \leq m}} (-1)^{k_3} \binom{n}{k_1}\binom{m}{k_2}\binom{m+1+k_3}{k_3} \frac{ q^{m-k_2}}{\left(q+1\right)^{m+k_3+2}}.
\end{multline}
In order to find the expression for the matrix elements of the momentum operator, we need to evaluate 2 integrals:
\begin{equation}
I_1(n,m)=\int_0^{\infty} dq\ (q+1)I(n+1,n,m,q),\quad I_2(n,m)=\int_0^{\infty} dq\ I(n,n,m,q).
\end{equation}
In order to perform the integrals, we notice that they can be expressed in terms of the beta function
\begin{equation}\label{eq:Beta_function}
B(x,y)=\int_0^{\infty} dt \frac{t^{x-1}}{(1+t)^{x+y}},\quad {\rm for\ }~ \Re{(x)}>0, ~\Re{y}>0.
\end{equation}
The first integral is:
\begin{equation}\label{eq:I_integrated}
I_1(n,m)=(-1)^n(\iu)^{n+1} \sum_{\substack{k_1, k_2, k_3 \\ k_1+k_2+k_3=n+1 \\ k_1 \leq n, k_2 \leq m}} (-1)^{k_3} \binom{n}{k_1}\binom{m}{k_2}\binom{m+1+k_3}{k_3} B(m-k_2+1,k_2+k_3).
\end{equation}
If $x,y\in \dN -\{0\}$, the beta function takes the form:
\begin{equation}
B(x,y)=\frac{(x-1)!(y-1)!}{(x+y-1)!}.
\end{equation}
Inserting this property into equation \eqref{eq:I_integrated} we get:
\begin{multline}
I_1(n,m)=(-1)^n \iu^{n+1} \sum_{\substack{k_1, k_2, k_3 \\ k_1+k_2+k_3=n+1 \\ k_1 \leq n, k_2 \leq m}} (-1)^{k_3} \binom{n}{k_1}\binom{m}{k_2}\binom{m+1+k_3}{k_3} \frac{(m-k_2)!(k_2+k_3-1)!}{(m+k_3)!} =\\=\frac{(-1)^n \iu^{n+1}}{(n+1)(m+1)} \sum_{\substack{k_1, k_2, k_3 \\ k_1+k_2+k_3=n+1 \\ k_1 \leq n, k_2 \leq m}} (-1)^{k_3} \frac{(n+1)!}{k_1! k_2! k_3!} \left(m+1+k_3 \right) =\\=\frac{(-1)^n \iu^{n+1}}{(n+1)(m+1)}\left(-(m+1)+ \sum_{\substack{k_1, k_2, k_3 \\ k_1+k_2+k_3=n+1 \\ k_1 \leq n+1, k_2 \leq m}} (-1)^{k_3} \frac{(n+1)!}{k_1! k_2! k_3!} \left(m+1+k_3 \right)  \right)\, .
\end{multline}
The second integral is:
\begin{multline}\label{eq:II_integrated}
I_2(n,m)=(-1)^n(\iu)^{n} \sum_{\substack{k_1, k_2, k_3 \\ k_1+k_2+k_3=n \\ k_1 \leq n, k_2 \leq m}} (-1)^{k_3} \binom{n}{k_1}\binom{m}{k_2}\binom{m+1+k_3}{k_3}  \frac{(m-k_2)!(k_2+k_3)!}{(m+k_3+1)!}=\\=\frac{(-1)^n(\iu)^{n}}{m+1} \sum_{\substack{k_1, k_2, k_3 \\ k_1+k_2+k_3=n \\ k_1 \leq n, k_2 \leq m}} (-1)^{k_3} \frac{n!}{k_1 ! k_2 ! k_3!} \, .
\end{multline}
The matrix elements of the momentum operator can be expressed in terms of the functions $I_1(n,m)$ and $I_2(n,m)$:
\begin{equation}
p_{nm}:=\sqrt{(n+1)(m+1)}\iu^{n+2}\left(-\frac{1}{2} I_1(n,m)-\iu I_2(n,m) \right)
\end{equation}

\subsubsection{Entries below and on the diagonal}
We will look for a compact formula for the matrix elements. To this end we will find expressions for
\be
S_{n,m}=\sum_{\substack{k_1, k_2, k_3 \\ k_1+k_2+k_3=n \\ k_1 \leq n, k_2 \leq m}} (-1)^{k_3} \frac{n!}{k_1 ! k_2 ! k_3!},\quad Z_{n,m}=\sum_{\substack{k_1, k_2, k_3 \\ k_1+k_2+k_3=n \\ k_1 \leq n, k_2 \leq m}} (-1)^{k_3} k_3 \frac{n!}{k_1 ! k_2 ! k_3!}.
\ee
We will find $S_{n,m}$ and $Z_{n,m}$ recursively. In this derivation we will rely on the following identity holding for any polynomial $P$ of degree smaller than $n$:
\begin{equation}
\sum_{k=0}^n (-1)^k \binom{n}{k} P(k)=0.
\end{equation}

Let us consider first $S_{n,m}$. From the definition, it is clear that
\begin{multline}
S_{n,m+1}=S_{n,m}+\sum_{\substack{k_1, k_2, k_3 \\ k_1+k_2+k_3=n \\ k_1 \leq n, k_2 = m+1}} (-1)^{k_3} \frac{n!}{k_1 ! k_2 ! k_3!}=\\=S_{n,m}+\frac{n!}{(n-m-1)!(m+1)!}\sum_{\substack{k_1, k_3 \\ k_1+k_3=n-m-1}} (-1)^{k_3} \frac{(n-m-1)!}{k_1! k_3!}=\\=\begin{cases}
S_{n,m}+1,&\textrm{ if }m=n-1,\\
S_{n,m},&\textrm{ if }m<n-1.\\
\end{cases}
\end{multline}
In order to find the desired identity, we notice that
\begin{equation}
S_{n,0}=\sum_{\substack{k_1, k_2, k_3 \\ k_1+k_2+k_3=n \\ k_1 \leq n, k_2 = 0}} (-1)^{k_3} \frac{n!}{k_1 ! k_2 ! k_3!}=\sum_{\substack{k_1, k_3 \\ k_1+k_3=n}} (-1)^{k_3} \frac{n!}{k_1 ! k_3!}=\delta_{n,0}.
\end{equation}
Combining the results we get that :
\begin{equation}
S_{n,m}=\delta_{n,m} \textrm{ for } m\leq n.
\end{equation}

The second sum will be calculated in completely similar manner. We have the following recurrence:
\begin{multline}
Z_{n,m+1}=Z_{n,m}+\sum_{\substack{k_1, k_2, k_3 \\ k_1+k_2+k_3=n \\ k_1 \leq n, k_2 = m+1}} (-1)^{k_3} k_3\, \frac{n!}{k_1 ! k_2 ! k_3!}=\\=Z_{n,m}+\binom{n}{m+1}\sum_{\substack{k_1, k_3 \\ k_1+k_3=n-m-1}} (-1)^{k_3} k_3 \binom{n-m-1}{k_3}=\\=\begin{cases}
Z_{n,m},&\textrm{ if }m=n-1,\\
Z_{n,m}-n,&\textrm{ if }m=n-2.\\
Z_{n,m},&\textrm{ if }m\leq n-3.
\end{cases}
\end{multline}
We can express the formula above using Kronecker delta:
\begin{equation}
Z_{n,m+1}=Z_{n,m}-n\delta_{n,m+2}.
\end{equation}
The recurrence starts at $Z_{n,0}$:
\begin{multline}
Z_{n,0}=\sum_{\substack{k_1, k_2, k_3 \\ k_1+k_2+k_3=n \\ k_1 \leq n, k_2 = 0}} (-1)^{k_3} k_3\, \frac{n!}{k_1 ! k_2 ! k_3!}=\sum_{\substack{k_1, k_3 \\ k_1+k_3=n}} (-1)^{k_3} k_3\, \binom{n}{k_3}=\begin{cases}
1,&\textrm{ if }n=0,\\
-1,&\textrm{ if }n=1.\\
0,&\textrm{ if }n\geq 2.
\end{cases}
\end{multline}
In our study, we will only need to consider the following two cases.
\begin{enumerate}
\item $n=1$. In this case $Z_{1,0}=-1$ and $Z_{1,1}= -1$.
\item $n\geq 2$. In this case $Z_{n,0}=0$ and
\begin{equation}
Z_{n,m+1}=Z_{n,m}-n\delta_{n,m+2}.
\end{equation}
As a result,
\begin{equation}
Z_{n,m}=-n\delta_{n,m+1}- n \delta_{n,m}.
\end{equation}
\end{enumerate}
Let us notice that the two cases can be written with one formula. For $n\geq 1$:
\begin{equation}
Z_{n,m}=-n\delta_{n,m+1}- n \delta_{n,m}.
\end{equation}

Let us now go back to the expression for $I_1(n,m)$ and $I_2(n,m)$.
\begin{multline}
I_1(n,m)=\frac{(-1)^n \iu^{n+1}}{(n+1)(m+1)}\left((m+1)(S_{n+1,m}-1)+ Z_{n+1,m} \right)=\\=\frac{(-1)^n \iu^{n+1}}{(n+1)(m+1)}\left((m+1)(\delta_{n+1,m}-1)-(n+1)\delta_{n+1,m+1} - (n+1\delta_{n+1,m} \right).
\end{multline}
The second object is:
\begin{equation}
I_2(n,m)=\frac{(-1)^n(\iu)^{n}}{m+1} S_{n,m}=\frac{(-1)^n(\iu)^{n}}{m+1} \delta_{n,m}.
\end{equation}
We will combine the results to obtain the expression for the matrix elements of the momentum operator:
\begin{equation}
p_{nm}=\frac{\iu}{\sqrt{(n+1)(m+1)}}\left(\frac{m+1}{2}(\delta_{n+1,m}-1)-\frac{n+1}{2}\delta_{n+1,m+1} - \frac{n+1}{2}\delta_{n+1,m}+(n+1) \delta_{n,m} \right).
\end{equation}
\begin{equation}
p_{nm}=\frac{\iu}{\sqrt{(n+1)(m+1)}}\left(-\frac{m+1}{2} + \frac{n+1}{2} \delta_{n,m} +\frac{m-n }{2}\delta_{n+1,m}\right).
\end{equation}
Let us notice that we make calculation for $m\leq n$ only. Therefore
\begin{equation}
p_{nm}=\frac{\iu}{\sqrt{(n+1)(m+1)}}\left(-\frac{m+1}{2} + \frac{n+1}{2} \delta_{n,m}\right).
\end{equation}
This means that:
\begin{equation}
p_{nn}=0,\quad p_{nm}=-\frac{\iu}{2}\sqrt{\frac{m+1}{n+1}}\textrm{ if }m < n.
\end{equation}

\subsubsection{The entries above the diagonal}

For a cross-check, we will calculate the entries above the diagonal as well $m>n$. In this case
\begin{equation}
S_{n,m}=\sum_{\substack{k_1, k_2, k_3 \\ k_1+k_2+k_3=n \\ k_1 \leq n, k_2 \leq m}} (-1)^{k_3} \frac{n!}{k_1 ! k_2 ! k_3!}=\sum_{\substack{k_1, k_2, k_3 \\ k_1+k_2+k_3=n \\ k_1 \leq n, k_2 \leq n}} (-1)^{k_3} \frac{n!}{k_1 ! k_2 ! k_3!}=(1+1-1)^n=1
\end{equation}
and
\begin{multline}
Z_{n,m}=\sum_{\substack{k_1, k_2, k_3 \\ k_1+k_2+k_3=n \\ k_1 \leq n, k_2 \leq m}} (-1)^{k_3}k_3\frac{n!}{k_1 ! k_2 ! k_3!}=\sum_{\substack{k_1, k_2, k_3 \\ k_1+k_2+k_3=n \\ k_1 \leq n, k_2 \leq n}} (-1)^{k_3} k_3 \frac{n!}{k_1 ! k_2 ! k_3!}=\\=x_3 \frac{\partial}{\partial x_3} (x_1+x_2-x_3)^{n}|_{x_1=x_2=x_3=1}=-n(1+1-1)^n=-n.
\end{multline}
Inserting the formulas into $I_1(n,m), I_2(n,m)$ and into $p_{nm}$ afterwards, we get:
\begin{equation}
p_{nm}=\frac{\iu}{2}\sqrt{\frac{n+1}{m+1}}\textrm{ if }m>n.
\end{equation}
This is consistent with the fact that the matrix $p$ is hermitian.

\subsubsection{Summary}

The calculations above show that
\begin{equation}\label{eq:momentum_operator}
p_{nn}=0,\quad p_{nm}=-\frac{\iu}{2}\sqrt{\frac{m+1}{n+1}}\textrm{ if }m < n,\quad  p_{nm}=\frac{\iu}{2}\sqrt{\frac{n+1}{m+1}}\textrm{ if }m>n.
\end{equation}

\subsection{The matrix elements of the coordinate operator}

The matrix elements of the coordinate operator:
\begin{equation}
q_{nm}:=\frac{\sqrt{(n+1)(m+1)}}{2\pi }\,\int_0^{\infty} dq \int_{-\infty}^{+\infty} dp\ \frac{q \left(\frac{q-1}{2} -\iu p \right)^{n}\left(\frac{q-1}{2} +\iu p \right)^{m}}{\left(\frac{q+1}{2} -\iu p \right)^{n+2}\left(\frac{q+1}{2} +\iu p \right)^{m+2}} .
\end{equation}
The integral over $p$ can be done using a contour method. We take the sunset contour in the  lower half plane and obtain
\begin{equation}
q_{nm}:=-\iu \sqrt{(n+1)(m+1)}\,\int_0^{\infty} dq\ q\,\Res{\frac{ \left(\frac{q-1}{2} -\iu p \right)^{n}\left(\frac{q-1}{2} +\iu p \right)^{m}}{\left(\frac{q+1}{2} -\iu p \right)^{n+2}\left(\frac{q+1}{2} +\iu p \right)^{m+2}}}{p=-\iu \frac{q+1}{2}} .
\end{equation}
We will calculate now the residuum:
\begin{multline}
\Res{\frac{ \left(\frac{q-1}{2} -\iu p \right)^{n}\left(\frac{q-1}{2} +\iu p \right)^{m}}{\left(\frac{q+1}{2} -\iu p \right)^{n+2}\left(\frac{q+1}{2} +\iu p \right)^{m+2}}}{p=-\iu \frac{q+1}{2}}=\\=\iu^{n+2} \frac{1}{(n+1)!} \frac{d^{n+1}}{d p^{n+1}} \left( {\frac{ \left(\frac{q-1}{2} -\iu p \right)^{n}\left(\frac{q-1}{2} +\iu p \right)^{m}}{\left(\frac{q+1}{2} +\iu p \right)^{m+2}}}\right).
\end{multline}
The integral that we will need to calculate this time is
\begin{equation}
I_3(n,m)=\int_0^{\infty} dq\, q\, I(n+1,n,m,q).
\end{equation}
With this integral, we can express the matrix elements of the coordinate operator in the following form:
\begin{equation}
q_{nm}=\iu^{n+1}\sqrt{(n+1)(m+1)} I_3(n,m).
\end{equation}
As previously, we can express the integral $I_3(n,m)$ as a sum.
\begin{multline}
I_3(n,m)=(-1)^n \iu^{n+1}\sum_{\substack{k_1, k_2, k_3 \\ k_1+k_2+k_3=n+1 \\ k_1 \leq n, k_2 \leq m}}  (-1)^{k_3} \binom{n}{k_1}\binom{m}{k_2}\binom{m+1+k_3}{k_3}\int_0^{\infty} dq\, \frac{q^{m-k_2+1}}{(q+1)^{m+k_3+2}}=\\=(-1)^n \iu^{n+1}\sum_{\substack{k_1, k_2, k_3 \\ k_1+k_2+k_3=n+1 \\ k_1 \leq n, k_2 \leq m}}  (-1)^{k_3} \binom{n}{k_1}\binom{m}{k_2}\binom{m+1+k_3}{k_3}\frac{(m-k_2+1)!(k_2+k_3-1)!}{(m+k_3+1)!}=\\
=-\frac{(-\iu)^{n+1}}{(n+1)(m+1)}\sum_{\substack{k_1, k_2, k_3 \\ k_1+k_2+k_3=n+1 \\ k_1 \leq n, k_2 \leq m}}  (-1)^{k_3} \frac{(n+1)!}{k_1! k_2! k_3!}(m+1-k_2)=\\=\frac{(-\iu)^{n+1}}{(n+1)(m+1)}\left((m+1)- \sum_{\substack{k_1, k_2, k_3 \\ k_1+k_2+k_3=n+1 \\ k_1 \leq n+1, k_2 \leq m}}  (-1)^{k_3} \frac{(n+1)!}{k_1! k_2! k_3!}(m+1-k_2)\right).
\end{multline}
This time we will consider separately the terms on the diagonal and above the diagonal (and we will not calculate the terms below diagonal.

\subsubsection{Diagonal terms}

\begin{multline}
q_{nn}=\frac{1}{n+1}\left((n+1)- \sum_{\substack{k_1, k_2, k_3 \\ k_1+k_2+k_3=n+1 \\ k_1 \leq n+1, k_2 \leq n}}  (-1)^{k_3} \frac{(n+1)!}{k_1! k_2! k_3!}(n+1-k_2)\right)=\\=\frac{1}{n+1}\left((n+1)- \sum_{\substack{k_1, k_2, k_3 \\ k_1+k_2+k_3=n+1 \\ k_1 \leq n+1, k_2 \leq n+1}}  (-1)^{k_3} \frac{(n+1)!}{k_1! k_2! k_3!}(n+1-k_2)\right)=\\=\frac{1}{n+1}\left((n+1)-(n+1)(1+1-1)^{n+1} + x_2 \frac{\partial}{\partial x_2}\left(x_1+x_2-x_3 \right)^{n+1}|_{x_1=x_2=x_3=1}\right)=1.
\end{multline}

\subsubsection{The terms above the diagonal}

We will consider matrix elements $q_{nm}$ for $m>n$. In this case:
\begin{multline}
q_{nm}=\frac{1}{\sqrt{(n+1)(m+1)}}\left((m+1)- \sum_{\substack{k_1, k_2, k_3 \\ k_1+k_2+k_3=n+1 \\ k_1 \leq n+1, k_2 \leq n+1}}  (-1)^{k_3} \frac{(n+1)!}{k_1! k_2! k_3!}(m+1-k_2)\right)=\\=\frac{1}{\sqrt{(n+1)(m+1)}}\left((m+1)-(m+1)(1+1-1)^{n+1}+x_2 \frac{\partial}{\partial x_2}\left(x_1+x_2-x_3 \right)^{n+1}|_{x_1=x_2=x_3=1}\right)=\\=\sqrt{\frac{n+1}{m+1}}.
\end{multline}

\subsubsection{Summary}
Since $q_{nm}$ is hermitian, we have
\begin{equation}\label{eq:coordinate_operator}
q_{nm}=\sqrt{\frac{n+1}{m+1}} \textrm{ if }m\geq n,\quad q_{nm}=\sqrt{\frac{m+1}{n+1}} \textrm{ if }m<n.
\end{equation}

\section{Evolution of quantum observables}\label{sc:evolution_of_obsevables}
In what follows, we consider the quantum evolution corresponding to the
classical case with $M < \frac{m}{2}$, where $m$ is the total rest mass of the
shell. Classically, the shell reaches the horizon $q_H = 2 M^2$ after an
infinite time (of an observer at spatial infinity), as it is illustrated in Fig. \!\ref{rysunek1}.

Remarkably, our quantum model experiences a different behaviour. In a
finite time, the quantum shell reaches a minimum size and after a bounce,
which is above the horizon, it expands until it reaches a maximum size.  More
precisely, in our quantum model we choose a state peaked at energy $M$.
One needs to remember that the energy of the system is conserved during the time evolution
given by the Schr\"{o}dinger equation.
We diagonalize the quantum Hamiltonian numerically. Due to technical limitations of our numerical
procedure we can study the model for small values of $M$ only. We calculate the quantum
evolution of the expectation values of the operators $\hat{q}$ and
$\hat{p}$. We notice that the expectation values $\langle q \rangle_t$ and
$\langle p \rangle_t$ are periodic functions of time $t$.

\subsection{Gaussian state}

We choose a state peaked at energy $M$ with standard deviation $\sigma$:
\begin{equation}
\ket{\Psi} = \frac{1}{N} \sum_i  e^{-\frac{(E_i-M)^2}{2\sigma^2}} \ket{E_i},
\end{equation}
where $\ket{E_i}$ is the eigenstate of the Hamiltonian operator with the
eigenvalue $E_i$ and $N$ is the normalization constant:
\begin{equation}
N^2=\sum_i e^{-\frac{(E_i-M)^2}{\sigma^2}}.
\end{equation}

In practical calculations we limit the sum to the region
$[M-4 \sigma\sqrt{2\ln(10)},M+4 \sigma\sqrt{2\ln(10)}]$.  With this choice, at
the end of the interval the exponential factor is
$\,\,-\frac{(E_i-M)^2}{2\sigma^2}=e^{-16\ln(10)}=10^{-16}$, which is
approximately equal to the machine precision for double accuracy used in our
calculations. We introduce a cut-off in the matrix size. We choose the parameter
$\sigma$ such that all eigenvalues in the region are good approximations of the
full eigenvalues.  We will choose eigenvalues which converge when the cut-off is
increased. The precise definitions will be given in the next section.

\subsection{Eigenvalues convergence}

In order to find the  Gaussian states and their evolution, we look for some
of the eigenvalues and eigenvectors of the Hamiltonian operator. We considered a
cut-off: $n< 1000,m< 1000$ and calculated the matrix elements $H_{nm}$ using
equation \eqref{eq:Hamiltonian_matrix}. We calculated the integrals numerically
using the DE rules for infinite range integrals as they are described in
\cite{NumericalRecipes}. This amounts to making a change of variables
\begin{equation}
q=e^{\pi \sinh(\tilde{q})},\quad p=\sinh(\pi \sinh(\tilde{p}))
\end{equation}
and performing the integrals in the range $]-4,4[$ using the open extended
trapezoidal rule (Trapz structure from \cite{NumericalRecipes}). We perform the
integral over $q$ first and afterwards we perform the integral over $p$.

\begin{figure}[!tbp]
  \includegraphics{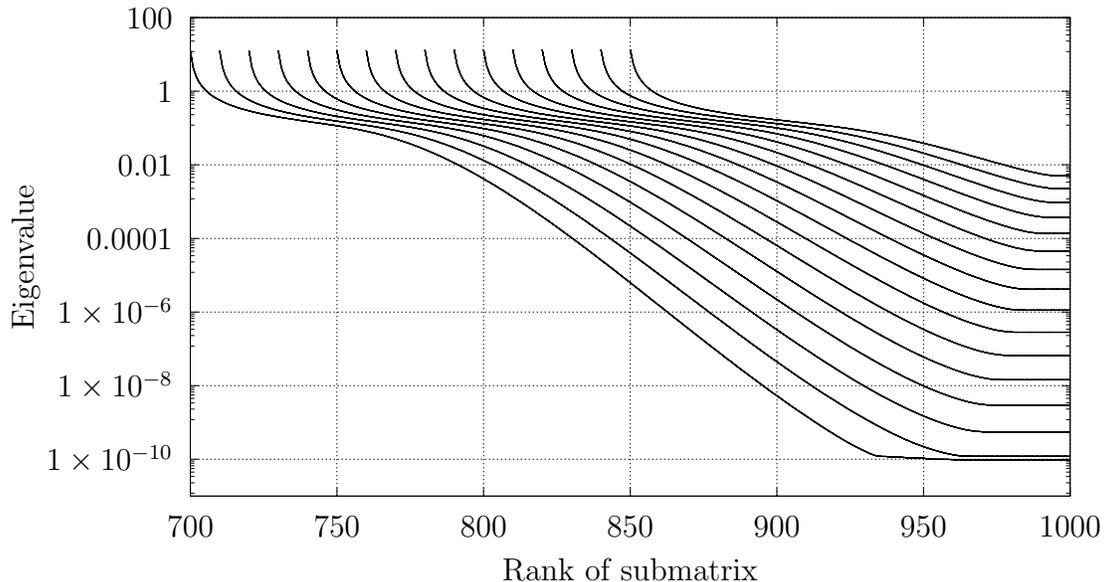}
  \caption{The eigenvalues $E_i(k), i=700,710,\ldots, 850$ of the submatrices of $H(k)$. The eigenvalue $E_i(k)$
  is the $i$-th eigenvalue of the sub-matrix $H(k)$ (the eigenvalues are sorted in the increasing order). Since
  the rank of $H(k)$ is equal $k$, it follows that $i\leq k$ and the plot of $E_i(k)$ starts at $k=i$. For each $i$
  we connected the values $E_i(k)$ with a line in order to guide the eye, i.e. each line represents one eigenvalue
  $E_i$.}\label{fig:eigenvalues_convergence}
\end{figure}

We look for convergence of the eigenvalues. Let us consider a family of
submatrices $H(k), k=1,\ldots,1000$, where each matrix $H(k)$ is obtained from
the Hamiltonian matrix $H$ by removing first $1000-k$ rows and first $1000-k$
columns. In particular $H(1000)=H$ and $k$ is the rank of the matrix $H(k)$. Let
us order the eigenvalues of $H(k)$ in the increasing order and let us denote by
$E_i(k), i\leq k$ the $i$-th eigenvalue of $H(k)$. We look for convergence of
the eigenvalues $E_i(k)$ as we increase $k$.  The plot
\ref{fig:eigenvalues_convergence} shows that eigenvalues $E_i$ for
$700\leq i \leq 850$ converge.  Each of the eigenvalues stabilizes, which is
reflected by the plateaux on the plot of $E_i$ as a function of $k$.

\subsection{Evolution of the observables}
We calculated the evolution of the expectation values of the operators $\hat{q}$ and $\hat{p}$ in a coherent state
peaked at energy $M=10^{-6}$ with standard deviation $\sigma=10^{-7}$. This means that we considered a state:
\begin{equation}
\ket{\Psi(t)} = \frac{1}{N} \sum_i  e^{-\frac{(E_i-M)^2}{2\sigma^2}-\iu\, E_i\, t}\, \ket{E_i}
\end{equation}
and calculated the expectation values
\begin{equation}
<\hat{q}>_t:=\scalar{\Psi(t)}{\hat{q}\,\Psi(t)}{\rm\ and\ }<\hat{p}>_t:=\scalar{\Psi(t)}{\hat{p}\,\Psi(t)}.
\end{equation}
The calculations are performed in the $e^{(1)}_n$ basis. After finding the eigenvectors of $\hat{H}$,
we calculated the components $\scalar{e^{(1)}_n}{\Psi(t)}$. Afterwards, we used the formulas
\eqref{eq:coordinate_operator} and \eqref{eq:momentum_operator} for the matrix elements of $\hat{q}$
and $\hat{p}$ in the $e^{(1)}_n$ basis.

\begin{figure}[!tbp]
  \centering
	{\includegraphics{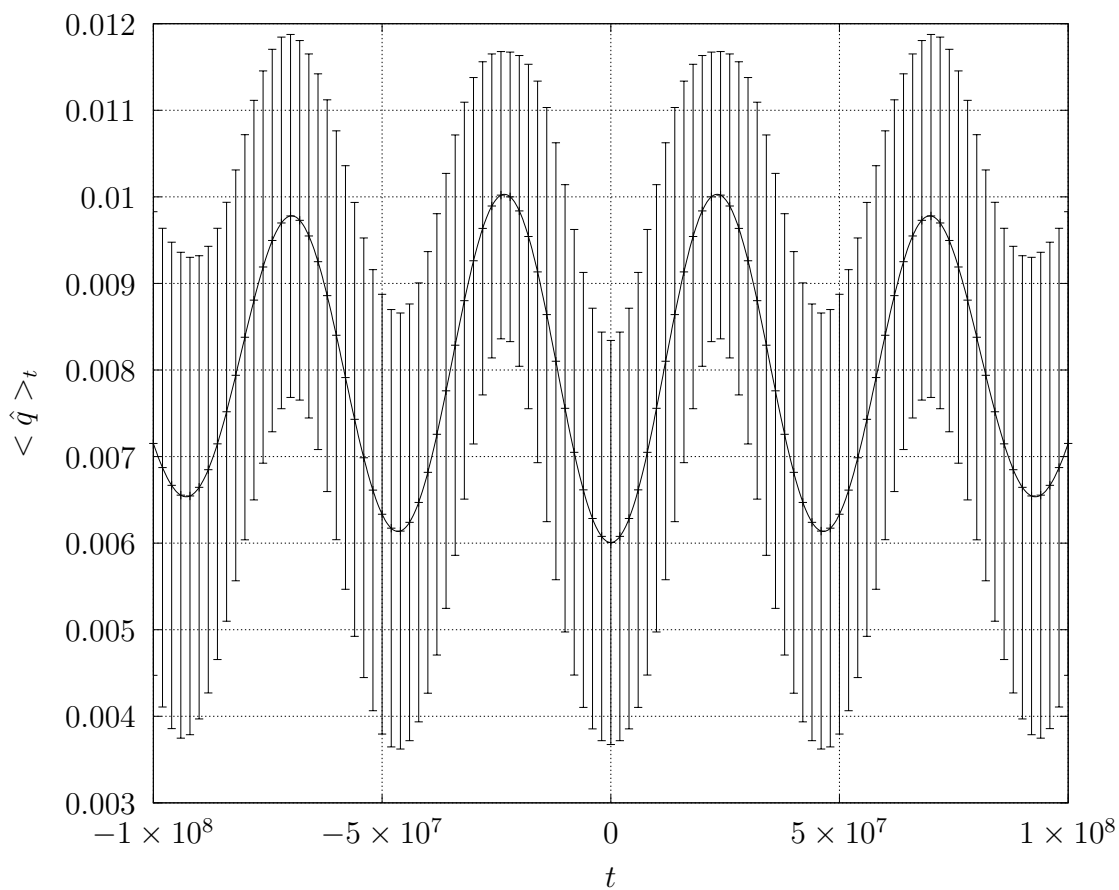}}
  \caption{Evolution of the expectation values of $\hat{q}$ in the coherent state peaked at energy $M=10^{-6}$ with standard deviation $\sigma=10^{-7}$. The expectation value is bounded from below by approximately $0.006$ ($0.037$ if the error bars are taken into account). This is well above the horizon which in this case is at $q_H=2M^2=2\cdot 10^{-12}$. We interpret this as a quantum bounce above the horizon. As a result, the evolution has oscillatory character.}\label{fig:q_evolution_of_expectation_values}
\end{figure}

We calculated $<\hat{q}>_t$ and $<\hat{p}>_t$ for $m=1.0, M=2\cdot 10^{-6}, \sigma=10^{-7}$.
The expectation value $<\hat{q}>_t$ is bounded from below
by approximately $0.006$. The bounce, connecting contracting and expanding branches, occurs well above
the horizon $q_H=2M^2 = 2\cdot 10^{-12}$ and at finite time as measured by an observer at spatial infinity.
We recall that in the corresponding classical case the shall falls onto the horizon at an infinite time.
The value $<\hat{q}>_t$ is also bounded from above by approximately $0.01$, which is smaller
than the classical maximal size of the shell $q_{\rm max}=\frac{m^4}{8(m-M)^2}\approx \frac{1}{8}$.
It is clear that $<\hat{q}>_t$ must oscillate as illustrated by Fig. \!\ref{fig:q_evolution_of_expectation_values}.

\begin{figure}[!tbp]
  \centering
	{\includegraphics{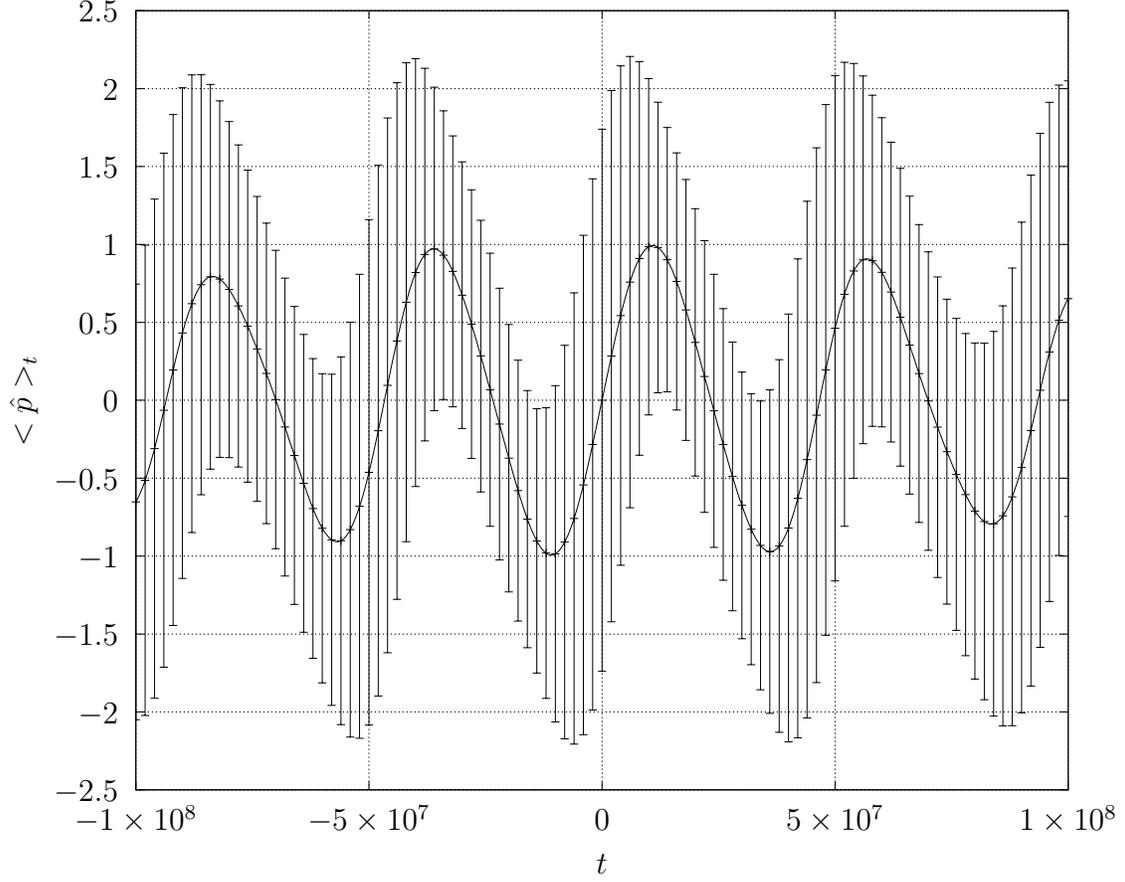}}
  \caption{Evolution of the expectation values of $\hat{p}$ in the coherent state peaked at energy $M=10^{-6}$ with standard deviation $\sigma=10^{-7}$. Let us notice that $<\hat{p}>_t=0$ when $<\hat{q}>_t$ is extremal. Classically this happened only when $q$ is maximal. }\label{fig:p_evolution_of_expectation_values}
\end{figure}

Since the classical variable $p$ is the measure of the angle between the surfaces of constant time on both sides
of the shell, it is reasonable to expect that the expectation value $<\hat{p}>_t$ oscillates as a consequence of the
oscillation of $<\hat{q}>_t$. Indeed, such an effect is seen in Fig. \!\ref{fig:p_evolution_of_expectation_values}.



\subsection{Oscillatory character of the evolution}\label{sc:periodic_evolution}
The oscillatory character of the evolution is related to our choice of the Gaussian state. This can be shown by the following calculation:
\begin{multline}\label{oscil}
<\hat{q}>_t=\frac{1}{N^2} \sum_{i,j} e^{-\frac{(E_i-M)^2}{2 \sigma ^2}-\frac{(E_j-M)^2}{2 \sigma ^2}} e^{\iu (E_i-E_j)t} \bra{E_i} \hat{q}
\ket{E_j}=\\=\frac{1}{N^2} \sum_{i,j} e^{-\frac{(E_i-M)^2}{2 \sigma ^2}-\frac{(E_j-M)^2}{2 \sigma ^2}} \cos( (E_i-E_j)t) \bra{E_i} \hat{q} \ket{E_j}.
\end{multline}
In the formula above, we used the fact that the eigenvectors $\ket{E_i}$ have real components in the $e^{(1)}_n$ basis and the fact
$q_{nm}=q_{mn}, ~q_{nm}\in \mathbb{R}$. Let us recall that in our numerical calculations we limited the sum to finite number of terms.
Therefore the resulting function is oscillatory as a finite sum of cosine functions.

Similar analysis applies to $<\hat{p}>_t$ giving trigonometric series.

We expect an oscillatory character if the spectrum is purely discrete. If the spectrum turns out to be continuous,
we expect that non-oscillatory functions could be achieved. In that case the sum in \eqref{oscil} is replaced by an integral
an the resulting function may be non-oscillatory. A typical example of such behaviour is a Fourier transform, which
decomposes non-oscillatory function into
oscillatory modes. However, full spectral analysis  is beyond the scope of this paper.

\section{Conclusions}
We used the coherent state quantization technique to investigate a quantum model
of a massive shell. We developed the new mathematical tools: we found a closed
formula for the basis elements $e^{(1)}_n$, and for the matrix elements of the
operators $\hat{q}$ and $\hat{p}$ in this basis. We studied the spectrum of the
quantum Hamiltonian operator numerically in the case $m=1.0$. We truncated the
operator spectrum and observed that some of the eigenvalues stabilize as we increase the
truncation.

 Using calculated eigenstates, we built a Gaussian state peaked at energy
$M=10^{-6}$.
We investigated
the evolution of the expectation values of $\hat{q}$ and $\hat{p}$ in the
Gaussian state and observed that the quantum shell bounces well above the
horizon. This leads to an oscillatory behaviour of the system which is in
contrast with the classical solution where the shell collapses and reaches the
Schwarzschild horizon $q_H=2M^2$ in infinite time (as measured by an observer at infinity).
The oscillatory character of the evolution may be traced back to our construction of the
Gaussian state. More detailed analysis of this phenomenon needs a careful study of the
spectral properties of the Hamiltonian operator.

Our analysis of the quantum Hamiltonian is based on  a series of
numerical experiments. The analytical analysis seems to be out of
range. According to these experiments it seems that this Hamiltonian, in the
investigated region, up to accuracy of our calculations (double accuracy)
has a continuous spectrum.

We realize that the research presented in this paper is not a complete
analysis. We developed the necessary tools and showed that this research
direction may lead to interesting results. In particular, the results concerning
oscillatory behaviour of the quantum system and the quantum origin of a bounce
above the horizon require independent cross-check. The integrals involved in the
expression of matrix elements of the Hamiltonian operator were calculationally
demanding. This resulted in rather limited size of the matrix and small value of
eigenvalues which can be reliably calculated. More efficient numerical
integration methods are needed in order to study the model in more details.
An advantage of our approach is that the quantum Hamiltonian  $\hat{H}$ we consider is
positive definite, which is generally true and independent on the chosen set of parameters.

The issue of the positivity of both classical and quantum Hamiltonians
  are of special interest.  At the classical level, it can be realized by
  restricting the phase space. The quantum level is more demanding.  In our
  integral quantization method there is a one-to-one correspondence between the
  phase space points and the space of coherent states. However, we cannot follow
  the classical procedure of choosing the subspace of all coherent states to
  obtain positivity of quantum Hamiltonian.  The reason is that such procedure
  would violate the mathematical consistency of the integral quantization and
  its physical interpretation as deformation of POV measure which results in
  problems with the probabilistic interpretation of this approach.  Therefore, in the integral
  defining the mapping of a classical observable into quantum observable one
  cannot restrict the region of integration, but one can restrict the integrant by
  taking $~\theta(H)H~$ instead of $H$.  This way one obtains the quantum
  Hamiltonian $\hat{H}$ which still acts on the full Hilbert space but it is
  positively defined because our quantization procedure maps positive functions
  on the phase space into positive operators.  Similar logic has been used, for
  instance, in the  paper \cite{Tomasz} within loop quantum gravity,
  where the dynamics is defined by a square root of positive part of quantum
  gravitational scalar constraint operator.

  To use this procedure consequently, the position and momentum quantum observables have
  to be defined on the whole Hilbert space. They determine physical meaning of
  the coherent states as quantum counterparts of the configuration space
  points. Our procedure is consistent with treating the operators $\hat{M}(Q)$, see
  Eq. \!(\ref{POVM}), as the POV measure.

\acknowledgments We would like to thank Jacek Jezierski, Jerzy Kijowski, and Daniele
Malafarina for helpful discussions. This work was partially supported by
the National Science Centre, Poland grant No. 2018/28/C/ST9/00157.

\appendix

\section{Orthonormal basis of the carrier space}
\label{basis}

The  basis of the Hilbert space $L^2(\dR_+, d\nu(x))$, where $d\nu(x):= dx/x$, is known  to be \cite{GM}
\begin{equation}\label{aa1}
e^{(\alpha)}_n (x) =
\sqrt{\frac{n!}{\Gamma (n + \alpha + 1)}}\,e^{-x/2} x^{(1 +\alpha)/2}\,L_n^{(\alpha)}(x),
\end{equation}
where $L_n^{(\alpha)}$ is the Laguerre function and  $\alpha > -1$.  One can verify that $\int_0^\infty
e^{(\alpha)}_n (x) e^{(\alpha)}_m (x) d\nu(x)= \delta_{n m}$ so that
$e^{(\alpha)}_n (x)$ is an orthonormal basis (for any fixed value of the parameter $\alpha > -1$).



\begin{thebibliography}{99}

\bibitem{KMM} J. Kijowski, G. Magli, and D. Malafarina, ``New derivation of the variational principle
for the dynamikcs of a gravitational spherical shell'', Phys. Rev. D {\bf 74}, 084017 (2006).

\bibitem{Vaz} C. Vaz, ``Proper time quantization of a thin shell'', arXiv:2205.06867 [gr-qc].

\bibitem{JK} J. Jezierski and J. Kijowski, ``Positivity of total energy in general relativity'',
Phys. Rev. D {\bf 36}, 1041 (1987).


\bibitem{Gozdz:2018aai}
  A.~G\'{o}\'{z}d\'{z}, W.~Piechocki and G.~Plewa,
  ``Quantum Belinski-Khalatnikov-Lifshitz scenario'',
  Eur.\ Phys.\ J.\ C {\bf 79}, 45 (2019).

\bibitem{AWT} A.~G\'{o}\'{z}d\'{z}, W. Piechocki, and T. Schmitz, ``Dependence of the affine
coherent states quantization on the parametrization of the affine group'', 	Eur. Phys. J. Plus {\bf 136}, 18 (2021).

\bibitem{Gel} I. M. Gel$'$fand and M. A. Na\"{i}mark, ``Unitary representations of the group of linear transformations of the
straight line'', Dokl. Akad. Nauk. SSSR {\bf 55}, 567 (1947).

\bibitem{AK1} E. W. Aslaksen and J. R. Klauder, ``Unitary
Representations of the Affine Group'', J. Math. Phys. {\bf 9}, 206
(1968).

\bibitem{AK2} E. W. Aslaksen and J. R. Klauder, ``Continuous Representation
Theory Using Unitary Affine Group'', J. Math. Phys. {\bf 10}, 2267 (1969).

\bibitem{BR} A. O. Barut and R. R\c{a}czka, {\it Theory of group representations and aplications} (PWN, Warszawa, 1977).


\bibitem{GM}  J.~P.~Gazeau and R.~Murenzi,
  ``Covariant affine integral quantization(s),''
  J.\ Math.\ Phys.\  {\bf 57},  052102 (2016).

\bibitem{Ber} H.~Bergeron and J.~P.~Gazeau,
``Integral quantizations with two basic examples,''
Annals Phys.\  {\bf 344}, 43 (2014).

\bibitem{Reed} M. Reed and B. Simon, {\it Methods of Modern Mathematical
    Physics} (San Diego, Academic Press, 1980), Vols I and II.

\bibitem{Arfken2011}  G. B. Arfken, H. J. Weber, F. E. Harris,
{\it Mathematical Methods for Physicists:  A Comprehensive Guide} (Academic Press, Oxford, 2011).

\bibitem{Busch1996}
P.~Busch, P.J.~Lahti and P.~Mittelstaedt, The Quantum Theory of Measurement,
second Rev. Edition,Springer-Verlag, Berlin Heidelberg, 1996,
ISBN 3-540-61355-2.

\bibitem{AOW} A. G\'{o}\'{z}d\'{z}, A. P\c{e}drak, and W. Piechocki,
``Ascribing quantum system to Schwarzschild spacetime with naked singularity'',
Class. Quantum Grav. {\bf 39}, 145005 (2022).



\bibitem{FaaDiBruno} F. Fa\`{a} di Bruno,
``Note sur une nouvelle formule du calcul diff\'{e}rentiel,'' Quart. J. Math. , 1 (1855) pp. 359-360.

\bibitem{BellPolynomialEncyclopedia} Bell polynomial. Encyclopedia of Mathematics.
URL:  \url{http://encyclopediaofmath.org/index.php?title=Bell_polynomial&oldid=46007}

\bibitem{BellPolynomial} E.T. Bell,
``Exponential polynomials,'' Ann. of Math. , 35 (1934) pp. 258-277.

\bibitem{Lah} I. Lah,
``Eine neue Art von Zahlen, ihre Eigenschaften und Anwendung in der mathematischen Statistik,'' Mitteil. Math. Statist. , 7 (1955) pp. 203-216

\bibitem{AbramowitzStegun} M. Abramowitz and I. A. Stegun, eds. {\it Handbook of mathematical functions with formulas, graphs,
and mathematical tables} (US Government printing office, 1964), Vol. 55.

\bibitem{NumericalRecipes}  W. H. Press,  S. A. Teukolsky, W. T. Vetterling, B. P. Flannery, Brian,
{\it Numerical Recipes:  The Art of Scientific Computing} (Cambridge University Press, 2007).

\bibitem{Tomasz} T. Paw{\l}owski and A. Ashtekar, ``Positive cosmological constant in loop quantum cosmology'',
Phys. Rev. D {\bf 85}, 064001 (2012).


\end{thebibliography}
\end{document}